# Brain Age from the Electroencephalogram of Sleep

Haoqi Sun[1], Luis Paixao[1], Jefferson T. Oliva[1,2], Balaji Goparaju[1], Diego Z. Carvalho[1], Kicky G. van Leeuwen[1,3], Oluwaseun Akeju[4], Robert Joseph Thomas[5], Sydney S. Cash[1], Matt T. Bianchi[1,*], M. Brandon Westover[1,*]

[1] Department of Neurology, Massachusetts General Hospital, Boston, MA, USA

[2] Bioinspired Computing Laboratory, Computer Science Department, University of São Paulo, Brazil

[3] University of Twente, Enschede, the Netherlands

[4] Department of Anesthesiology, Critical Care and Pain Medicine, Massachusetts General Hospital, Boston, MA, USA

[5] Department of Medicine, Division of Pulmonary, Critical Care & Sleep, Beth Israel Deaconess Medical Center, Boston, MA, USA

# Abstract

The human electroencephalogram (EEG) of sleep undergoes profound changes with age. These changes can be conceptualized as "brain age", which can be compared to an age norm to reflect the deviation from normal aging process. Here, we develop an interpretable machine learning model to predict brain age based on two large sleep EEG datasets: the Massachusetts General Hospital sleep lab dataset (MGH, N = 2,621) covering age 18 to 80; and the Sleep Hearth Health Study (SHHS, N = 3,520) covering age 40 to 80. The model obtains a mean absolute deviation of 8.1 years between brain age and chronological age in the healthy participants in the MGH dataset. As validation, we analyze a subset of SHHS containing longitudinal EEGs 5 years apart, which shows a 5.5 years difference in brain age. Participants with neurological and psychiatric diseases, as well as diabetes and hypertension medications show an older brain age compared to chronological age. The findings raise the prospect of using sleep EEG as a biomarker for healthy brain aging.

# Introduction

Human sleep undergoes profound changes with age, reflected in both the overall sleep architecture and electroencephalogram (EEG) oscillations/waveforms. At the level of sleep architecture (macro-structure), older participants tend to sleep earlier and awake earlier; have shorter sleep duration, increased sleep fragmentation and reduced percentages of rapid eye movement (REM), as well as (at least in males) deep non-REM (NREM) sleep (Mander et al., 2017; Scullin, 2017). At the level of EEG microstructure, older participants exhibit reduced slow waves during deep sleep (Carrier et al., 2001; Larsen et al., 1995), decreased sleep spindle amplitude, density and duration (Purcell et al., 2017), and less phase coupling between slow oscillations and sleep spindles (Helfrich et al., 2017). However, no study has yet addressed the inverse problem: how accurately can a person's age be predicted from the sleep EEG? What factors make a person's "brain age" (BA) older or younger than its chronological age (CA)?

Beyond predicting age from sleep EEG, brain age serves as a potential aging biomarker where variation between individuals at the same age may carry important information about the risk of cognitive impairment, neurological or psychiatric disease and death. Various biomarkers of aging have been proposed to better predict life span and functional capability other than EEG, ranging from molecular and cellular level to organ-system level. From the molecular and cellular aspects of aging, related biomarkers include leucocyte telomere length (Kruk et al., 1995), Ink4/Arf locus expression (Krishnamurthy et al., 2004), N-glycan profile (Vanhooren et al., 2010), mitochondrial DNA deletions (Eshaghian et al., 2006), and DNA-methylation status at CpG sites across the genome (Horvath, 2013), among others. From the organ-systems level, brain anatomy changes dramatically throughout life (Raz and Rodrigue, 2006). For example, cortical volume (Cole, 2017), thalamic volume (Redline et al., 2004), and white matter integrity (Mander et al., 2017), each decrease with aging. Using magnetic resonance images (MRI), chronological age can be predicted with mean absolute deviation of 5 years in healthy participants (Franke et al., 2010). Various diseases, including Alzheimer's disease, schizophrenia, epilepsy, traumatic brain injury, bipolar disorder, major depression, cognitive impairment, diabetes mellitus, and HIV, are associated with older brain age than chronological age (Cole et al., 2017; Cole,

2017; Cole and Franke, 2017; Cole et al., 2016; Franke et al., 2012; Franke et al., 2010). Alongside these biomarkers, EEG-based brain age is a sensible potential complement, which has several advantages: (1) EEG-based brain age could reflect functional changes rather than anatomical changes; (2) EEG is more economical, participant friendly, and in principle could be measured by home-based devices; (3) EEG-based brain age could facilitate within-participant repeated measures to assess the effectiveness of interventions, such as medications (Roehrs and Roth, 2010) or brain stimulation (Tasali et al., 2008) that aim to preserve or improve brain function.

We developed an EEG-based brain age estimate based on two sleep EEG datasets: the Massachusetts General Hospital sleep lab (MGH) dataset (Biswal et al., 2017; Sun et al., 2017), and the Sleep Hearth Health Study (SHHS) (Dean et al., 2016; Quan et al., 1997; Redline et al., 1998). We identified participants with neurological or psychiatric diseases in the MGH dataset, and trained the model only on participants without these diseases so that the scatter around the line of identity in testing participants is less affected by the clinical heterogeneity in the datasets. The effects of different EEG electrode choices were also investigated. The model was then validated on a longitudinal study in the SHHS dataset. We correlated the deviation of brain age from chronological age with clinical covariates, diseases and medications. Finally, an age norm was derived based on sleep EEG features, using which we can interpret the presence of deviations from chronological age.

## Material and Methods

### Dataset

The Partners Institutional Review Board approved retrospective analysis of the de-identified polysomnograms (PSG) dataset, acquired in MGH between 2008 and 2012, without requiring additional consent for its use in this study. The EEG signals contain 6 channels: frontal (F3-M2 and F4-M1), central (C3-M2 and C4-M1) and occipital (O1-M2 and O2-M1), each referenced to the contralateral mastoid. EEG signals are sampled at 200Hz. The signals are segmented into non-overlapping 30-second epochs. There are seven technicians who scored the entire PSGs according to AASM standards (AASM, 2007) into one of five stages: wake (W), rapid eye movement (REM), Non-REM stage 1 (N1), Non-REM stage 2 (N2) and Non-REM stage 3 (N3). Each PSG is scored by one technician. The inclusion criteria include (1) age between 18 to 80 years; (2) diagnostic PSG; and (3) with clinical diagnosis not later than one year after the PSG recording. We exclude EEGs without all 5 sleep stages. We investigate the effect of imputation of missing sleep stages in the supplementary material (Figure S2).

We define a participant having "neurological or psychiatric disease" as having at least one neurological or psychiatric disease in Table S1 in supplementary material, not later than one year after the PSG recording. This definition is consistent with the one used in Brain Images of Normal Participants (BRAINS) (Job et al., 2017), Open Access Series of Imaging Studies (OASIS) (Marcus et al., 2010), and in line with (Steffener et al., 2016) and (Cole et al., 2015). We refer participants without any neurological or psychiatric disease as "healthy".

In total, we identify 2,621 EEGs where 189 of them have neurological or psychiatric diseases. Table 1 provides summary statistics for the dataset.

**Table 1.** MGH Sleep Dataset Summary

| Characteristics | Value: Median (IQR) |
|---|---|
| Number of EEGs | 2,621 (including 189 EEGs with neurological or psychiatric diseases*) |
| Age (year) | 50 (37 – 61) |
| Gender | Female 1,314 (50%), Male 1,307 (50%) |
| BMI (kg/m$^2$) | 29 (25 – 35) |
| Overall AHI (per hour) | 5 (2 – 12) |
|     Normal (<5) | 1,273 |
|     Mild sleep apnea (5<=AHI<15) | 846 |
|     Moderate sleep apnea (15<=AHI<30) | 383 |
|     Severe sleep apnea (AHI>=30) | 119 |
| Medication | |
|     No medication | 682 |
|     Systemic | 1,493 |
|     Sleep aid | 823 |
|     Hypertension | 755 |
|     Antidepressant | 661 |
|     Neuroactive | 509 |
|     Benzodiazepine | 465 |
|     RLS/PLMS drugs | 242 |
|     Diabetic | 211 |
|     Opiate | 195 |
|     Antihistamine | 151 |
|     Z-drugs | 143 |
|     Herbal | 126 |
|     Stimulant | 121 |
|     Neuroleptic | 101 |

**Legend**: IQR: interquartile range; BMI: body mass index; AHI: apnea-hypopnea index (number of apnea events per hour of sleep) at 4% oxygen desaturation for hypopnea; *: Refer to Table S1 in the supplementary material for the detailed list of diseases that are counted as neurological or psychiatric.

We also use a subset of the SHHS dataset (Dean et al., 2016; Quan et al., 1997; Redline et al., 1998), which contains repeated EEGs from the same participant from two visits about 5 years apart, making it possible to evaluate the longitudinal within-participant reliability of our model. Selection criteria are: (1) having EEGs from both visits; (2) having a chronological age between 40 to 80 years in both visits (minimum age in SHHS is 40); and (3) having EEG and sleep stage scoring of high quality according to SHHS specifications. We also exclude EEGs without all 5 sleep stages. As a result, 1,760 EEGs from SHHS visit 1 and 1,760 paired EEGs from visit 2 are used. 24 participants have EEGs 4 years apart, 1,294 participants 5 years apart, 434 participants 6 years apart, and 7 participants 7 years apart. Unlike the MGH dataset, the SHHS dataset uses two EEG channels (C3-M2 and C4-M1) and is scored according to R&K standard (Rechtschaffen, 1968). To make scoring consistent across datasets, we combine S3 and S4 in SHHS to match the scoring of N3 in the MGH data. Summary statistics for the SHHS dataset are shown in Table 2.

**Table 2.** Sleep Heart Health Study Dataset Summary

| Characteristics | SHHS visit 1 (SHHS1) Median (IQR) | SHHS visit 2 (SHHS2) Median (IQR) |
|---|---|---|
| Number of EEGs | 1,760 | 1,760 (same participants as SHHS1) |
| Age (year) | 60 (54 – 67) | 65 (59 – 73) |
| Gender | Female 965 (55%), Male 794 (45%) | |
| BMI (kg/m$^2$) | 28 (25 – 31) | 28 (25 – 31) |
| Overall AHI (per hour) | 3 (1 – 9) | 6 (2 — 13) |

### EEG Preprocessing and Artifact Removal

EEG signals are notch-filtered at 60Hz to reduce line noise, and bandpass filtered from 0.5Hz to 20Hz to reduce myogenic artifacts. For 30s-epochs, those with absolute amplitude larger than 500uV are removed to restrict movement artifacts. Epochs containing flat EEG for more than 2 seconds are also removed. To reduce inter-participant variance, the amplitude of each EEG channel is normalized to have zero median and unit interquartile range across the whole night. The total amount of data removed by these preprocessing procedures is 7% in the MGH dataset and 9% in the SHHS dataset.

### Feature Extraction

For brain age prediction, we extracted features from the EEG used for sleep staging in our previous work (Sun et al., 2017). We extract 102 features from each 30-second epoch covering both time and frequency domains, as summarized in Table S2. For each EEG recording, we average the features in each of the 5 sleep stages over time, yielding 102 × 5 = 510 features per EEG. The features are log-transformed to render the feature distributions approximately Gaussian. Then the features are then z-transformed to have zero mean and unit standard deviation in the training set, and applied to the testing set.

### Brain Age Prediction

The model minimizes an objective function $J(w, b)$ with two terms: (1) the mean squared prediction error; and (2) the magnitude of the covariance between $CA$ and $DA$ ($DA = BA - CA$):

$$J(w, b) = \langle DA^2 \rangle + \lambda |Cov(CA, DA)|$$

where $BA_i = \text{softplus}(w^T x_i + b) = \ln\left[1 + e^{w^T x_i + b}\right]$, i.e. a linear combination of EEG features followed by a softplus function to ensure positivity of $BA$; $\langle DA^2 \rangle = \sum_{i=1}^{N}(DA_i)^2/N = \sum_{i=1}^{N}(BA_i - CA_i)^2/N$ is the mean squared prediction error; and $Cov(CA, DA) = \sum_{i=1}^{N}|(CA_i - \langle CA \rangle)(DA_i - \langle DA \rangle)|$ is the average absolute correlation (covariance) between the chronological age $CA_i$ and brain age deviation $DA_i$. Minimizing the first component of the objective function $\langle DA^2 \rangle$ encourages the model to produce predictions BA that are accurate (close to CA). Minimizing the second component $Cov(CA, DA)$ minimizes systematic prediction error, i.e. encourages deviations of BA from CA to be uncorrelated with CA. The second component is weighted by a hyperparameter λ, which is tuned to optimize the tradeoff between the two components.

To determine the optimal hyperparameter λ, we randomly select 300 EEGs from the training set to serve as internal validation data. We perform grid search on λ ranging from [0, 1, 5, 10]. We train the model on the remaining training EEGs and find the hyperparameter that has the highest performance on the internal validation set. The performance metric for each hyperparameter is evaluated using $\text{corr}(CA, BA) - |\text{corr}(CA, DA)|$, where corr(·) denotes Pearson's correlation. A larger $\text{corr}(CA, BA)$ and a smaller $|\text{corr}(CA, DA)|$ indicate better model performance. Here the metric is correlation instead of covariance or prediction error, since correlation is normalized and comparable among different hyperparameter values. Once the optimal hyperparameter with the best validation performance are determined, the validation set is combined with the rest of the training set, and the model is re-trained using the optimal hyperparameter.

We maintain strict separation of training and testing data to achieve statistically unbiased estimates of model's performance. The reported results are based on the testing set if not specifically stated.

## Results

**Predicting Brain Age from EEG**

The MGH dataset is randomly partitioned into a training set (N = 1,343) used to train the model where all of them are healthy, and a testing set (N = 1,278) to evaluate model performance among which 189 have neurological or psychiatric diseases and the rest 1,089 don't. The comparison between these two groups is presented in the later subsection. Here we first compare the age prediction performance between using sleep macro-structure features (Mander et al., 2017; Redline et al., 2004) (30 features in total, Table S3), and using the 510 sleep micro-structure EEG features, on the testing set. As shown in Figure 1, the mean absolute deviation (MAD) of BA from CA using macro-structure features is much higher (23.5 years) than that using micro-structure features (8.5 years), and the correlation is weaker (0.44, 95% CI 0.40 – 0.48) than that using micro-structure features (0.79, 95% CI 0.77 – 0.81). The results suggest that the effects of age on brain activity are more consistently reflected in sleep EEG micro-structure than in sleep stage composition (macro-structure). The progression of EEG features with age is further illustrated by a two-dimensional visualization of the feature space in Figure S1.

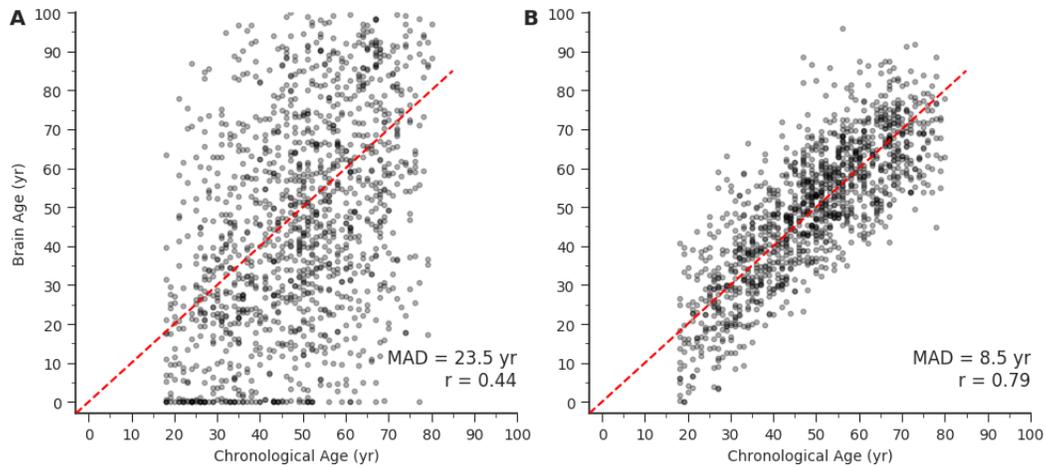

**Figure 1.** Scatterplot of predicted BA vs. CA using (A) sleep EEG macro-structure features and (B) sleep micro-structure features (Table S3). Both plots are obtained from the same testing participants. The red dashed diagonal line is the identity line where BA is equal to CA. The training set contains only healthy participants, while the testing set shown here contains participants both with and without neurological or psychiatric diseases.

In Figure 2, we show spectrograms of sleep EEGs from 6 typical individuals, arranged as a "confusion matrix", where rows show participants with CA in young (18 – 30 years), intermediate (40 – 50 years), and older adult groups (70 – 80 years) from top to bottom, and columns with BA in three groups from left to right. The participants positioned along the diagonal line are examples of participants with matched CA and BA in three groups. As an example, the top middle panel indicated by * shows a 25-year-old female who has been accumulatively diagnosed not later than one year after the study with hypertenstive disorder, post-traumatic stress disorder, anxiety, bipolar disorder, dizziness, hypercholesterolemia, asthma, irregular periods and. Her EEG shows strong and lasting theta power compared to a healthy participant at similar age of 26 (top left), which resembles an older BA at 40s (middle column).

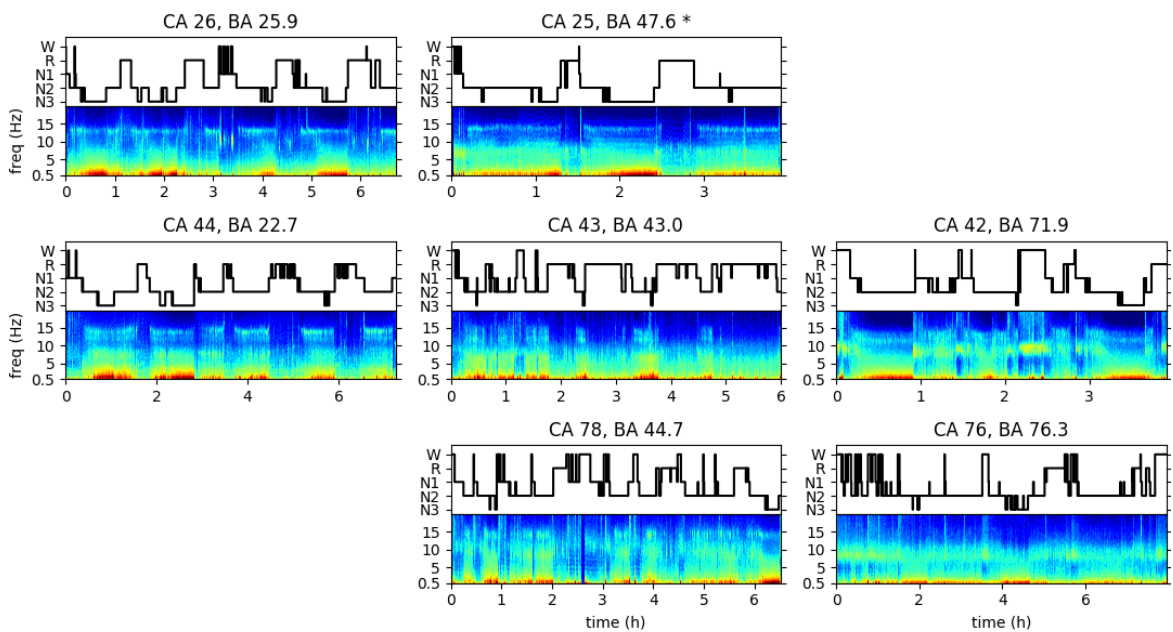

**Figure 2.** Individual cases with BA matches (diagonal) or is younger or older than CA (off diagonal). Each panel consists of a hypnogram and the corresponding spectrogram. Spectrograms are calculated as the average across the 6 EEG channels. The horizontal axis is time in hours. The case indicated by * is described in the main text.

**Effect of EEG Electrode Placement**

Most home-based EEG devices use fewer electrodes than lab-based PSG. We therefore explore how the EEG-based age prediction depends on the number and location of EEG electrodes. The prediction performance on the 1,089 healthy testing participants using different subsets of EEG electrodes is shown in Table 3. Using all six electrodes (2 frontal, 2 central 2 occipital) provides the lowest prediction error (8.1 years) and the highest correlation with CA (0.81) (Kruskal-Wallis p-value < 0.01). If limited to recording EEG from one brain area only, the frontal or occipital electrode provide similar performance (p-value = 1), while being significantly better than the central electrode (p-value < 0.01). Therefore it is suggested to use frontal EEG electrodes, if required to use less EEG electrodes.

**Table 3.** Brain age using subsets of EEG electrodes in healthy testing participants

| Electrode | Mean absolute deviation (years) | Pearson's correlation |
|---|---|---|
| Frontal (F3 and F4) | 10.1 | 0.78 |
| Central (C3 and C4) | 12.5 | 0.76 |
| Occipital (O1 and O2) | 10.2 | 0.75 |
| All electrodes | 8.1 | 0.81 |

**Longitudinal Reliability of the Brain Age Model**

To assess the longitudinal reliability of the brain age model, we use a subset of the SHHS dataset where each individual underwent two study visits, referred as SHHS1 and SHHS2 (1,760 adults, for a total of 3,520 nights of EEG). The average time difference between the two visits is 5.2 years (SD 0.5 year) (Table 2). We train the model in two ways. First, we train the model on 935 participants with paired EEGs (1,870 EEGs) from both visits and test the held out 825 participants with paired EEGs (1,650 EEGs) in both visits. The testing results are shown in Figure 3A and 3B. The MAD is 8.5 years on SHHS1 and 7.7 years on SHHS2, and 8.1 years overall all. The Pearson's correlation is 0.63 (95% CI 0.59 – 0.67) on SHHS1 and 0.66 (95% CI 0.62 – 0.70) on SHHS2. The average difference in predicted BA between the two visits (BA2 – BA1) is 5.5 years (SD 8.9 years) (Figure 3C). A t-test reveals no significant difference between BA2 – BA1 and CA2- CA1 (p-value = 0.3).

Second, we train the model on the 2,432 EEGs from the healthy participants in MGH dataset using the C3-M2 and C4-M1 only (to be consistent with the SHHS dataset), and then predict BA on the 3,520 paired EEGs in SHHS as the testing set. The results on the testing set are shown in Figure 3D and 3E. The MAD is 8.8 years on SHHS1 and 9.2 years on SHHS2. The Pearson's correlation is 0.61 (95% CI 0.57 – 0.64) on SHHS1 and 0.56 (95% CI 0.52 – 0.60) on SHHS2. The average difference in predicted BA between the two visits (BA2 – BA1) is 4.3 years (SD 10.3 years) (Figure 3F). A t-test reveals a significant difference between BA2 – BA1 and CA2- CA1 (p-value < 0.01).

The results indicate that the model is able to accurately track the 5-years of aging elapsed between study visits at the population level when trained on part of SHHS and tested on the other part. When trained on MGH and tested on SHHS the model shows good generalization for age 40 to 70. The performance is modestly reduced as limited by having two central EEG electrodes in SHHS dataset instead of six in MGH dataset.

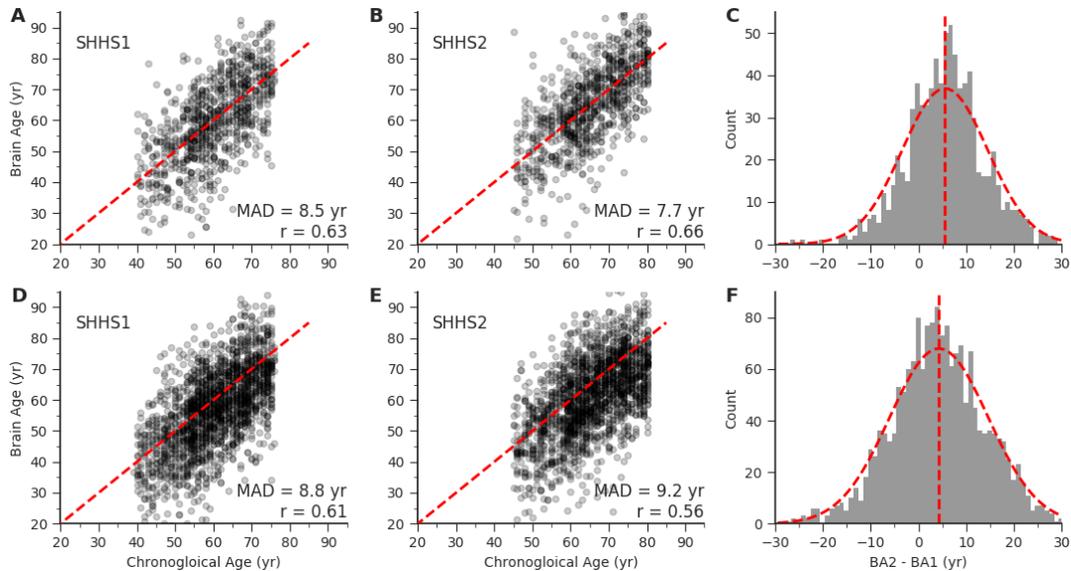

**Figure 3**. (A and B) CA vs. BA from the 1,650 testing EEGs in SHHS when trained on the other part of SHHS. (D and E) CA vs. BA from the 3,520 testing EEGs in SHHS when trained on the healthy participants in MGH dataset. (C and F) Histogram of BA differences between two SHHS visits, showing tracking of age at the population level. The dashed red line is the Gaussian fit to the histogram.

**Correlation between Covariates and DA (BA – CA)**

We collect various covariates from the MGH dataset, including sleep macro-structure (i.e. sleep stages), the Epworth Sleepiness Scale, and measures of sleep disruptions and obtain the Spearman's correlation with the deviation of EEG-based brain age from CA (DA = BA - CA). For the healthy participants, the covariates with statistically significant Spearman's correlation are shown in ascending order in Table 4. The full list of all covariates is shown in Table S4 in the supplementary material. The amount of deep sleep shows the most negative correlation with DA, although being weak. Wake time after sleep onset also shows weak but significant positive correlation with DA.

The weak correlations indicate that DA is relatively independent from these covariates, such as apnea-hypopnea index (AHI). Therefore DA can be interpreted as a relatively orthogonal metric for measuring sleep and brain health. In other words, DA is robust in participants with different characteristics. As an example, a scatter plot between AHI and CA/BA/DA shown in Figure S3 in the supplementary material reveals an correlation between AHI and CA/BA, not but DA, since it is canceled by taking the difference.

**Table 4**. Covariates that have significant correlation with DA (BA-CA)

| Covariate | Spearman's correlation |
|---|---|
| N3 time | -0.12 |
| N3 percentage | -0.097 |
| Sleep efficiency | -0.075 |
| Total sleep time | -0.065 |
| N2 time | -0.058 |
| Respiratory Disturbance Index | 0.044 |
| N1 percentage | 0.054 |
| Wake Time After Sleep Onset | 0.078 |

**Correlation between Disease / Medication and DA (BA – CA)**

We next show DA in different health conditions. In Section Material and Methods we have described the criteria of being considered as having neurological or psychiatric diseases (Table S1), which are hypothesized to have older BA than CA. This hypothesis is verified as shown in Figure 4A. The DA for healthy group (2,432 participants) is around 0 year, while the DA for the group with neurological or psychiatric diseases (189 participants) increases up to 5 years with a significant difference (t-test p-value < 0.01). Therefore on a population level, the participants with neurological or psychiatric diseases have an average 5 years older brain age than their chronological age. Further, in the healthy group, the MAD is 8.1 years and correlation between CA and BA is 0.81; in the group with neurological or psychiatric diseases, the MAD is larger at 10.6 years and correlation smaller at 0.70, both of which are significant (p-value < 0.01).

Despite grouping participants based on their diagnosis of diseases, it is also worthy looking at how the medication affects DA. The MGH database includes self-reported types of medications in each of the 14 gross categories prescribed at the time of EEG recording. The exact type or the dose is currently not provided as limited by the dataset. We search participants for all possible combinations of taking 1 to 4 categories of medications. For each combination of medication categories, we apply a t-test to compare the DA of two groups: participants not taking any of the medications in this combination vs. participants taking all medications in this combination only (not taking others). Groups with less than 20 participants are excluded from the analysis to ensure statistical validity. The medication categories that have significantly different DA between the groups (t-test p-value < 0.05) are listed in Table 5.

**Table 5.** DA (BA - CA) under different medication categories that is significant

| Medication Category | DA without medications: mean (std) in years | DA with medications: mean (std) in years | #participant without medications | #participant with medications |
|---|---|---|---|---|
| Diabetic + Hypertension | -0.49 (9.7) | 3.8 (9.7) | 1671 | 21 |
| Hypertension + Systemic | -0.68 (9.6) | 1.2 (10.2) | 951 | 331 |

We hypothesize that participants with diabetics and hypertension tend to have older BA than CA. As shown in Figure 4B, participants taking medications for diabetics and hypertensions have significantly larger DA compared to those who are not taking these medications (p-value < 0.05). Comparison to the age norm reveals that the reason is that these participants tend to have more continual theta power during N1 and N2 stages. For benzodiazepine medications, the model gives younger brain age because of the they promotes the spindles which appears to be younger from the model point of view; similarly for opiate medications, the model also gives young brain age because they slows down EEG and hence more delta power and appears to be younger from the model point of view.

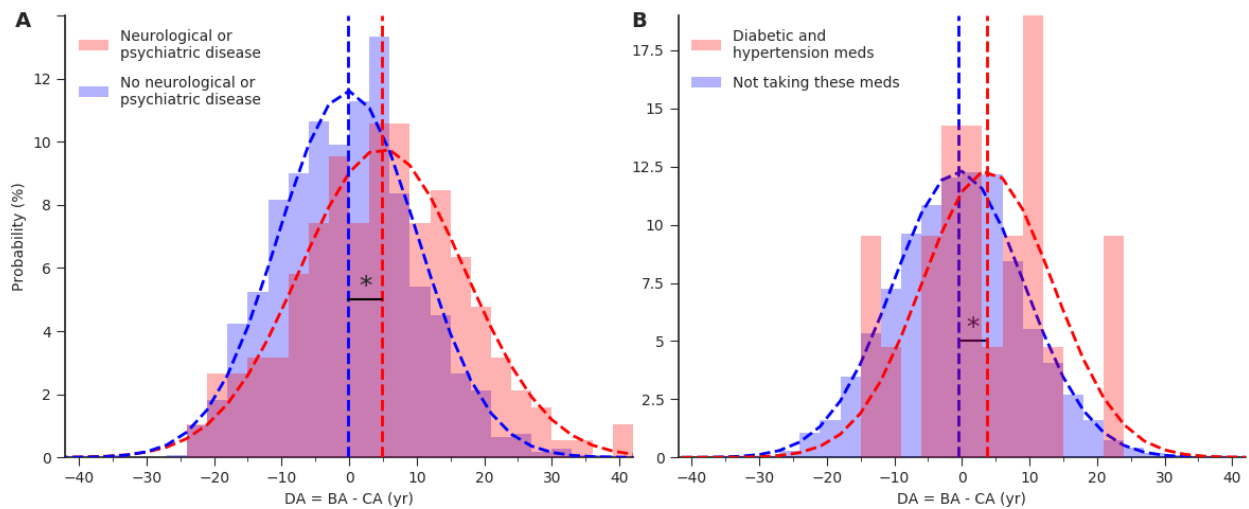

**Figure 4.** (A) Histograms of DAs for participants with (red) and without (blue) neurological or psychiatric diseases defined in Table S1. The dashed lines show the Gaussian fit to these distributions. The difference in the mean value in the two groups is 5.1 years. (B) Histograms of DAs for participants taking (red) and not taking (blue) medications for diabetics and hypertension. The different in the mean value in the two groups is 4.3 years.

**Age Norm based on Sleep EEG Features**

The parametric formulation of our brain age model allows studying individual differences at the level of EEG features, as contrast to nonparametric models such as Gaussian process regression. To do this, we first derive the "age norm", i.e. typical EEG feature values at different CAs with matched BAs. Specifically, the age norm of age x is obtained by averaging the z-transformed EEG features from all healthy participants who have both CA and BA within [x-5, x+5] years, as shown in Figure 5.

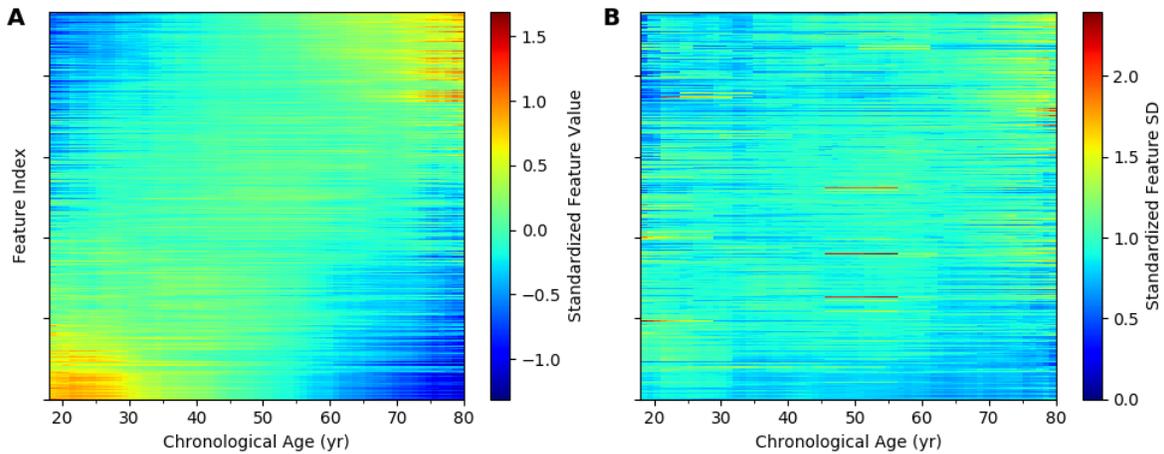

**Figure 5**. The age norm represented as a heat map. The features are ranked based on their correlation with CA. Panel A shows the age norm, i.e. average z-transformed EEG features. Panel B shows the standard deviation of the age norm.

The age norm suggests that when following a normal aging process, the features with most negative correlation with age is delta band (< 4Hz) power in N3 stage in occipital and central areas; and the features with most positive correlation with age is theta band (4 – 8Hz) power in N1 stage in frontal, occipital and central areas (Table S5).

Using this feature age norm, we can examine the reasons of the deviation from CA for the exemplar participants in Figure 6 instead of visual inspection of the spectrograms. This is a 29-year-old female accumulatively diagnosed as diabetes mellitus, obesity, smoker, cerebrovascular accident, congestive heart failure and acute renal failure not later than 1 year after the PSG recording. Her brain age is as high as 63-year-old. When compared to the age norm, her sleep EEG exhibits (1) more theta band bursts compared to the normal REM spectrum which has low theta band activity; (2) less delta to theta ratio during deep sleep; and (3) less theta and alpha band bursts, i.e. more continual theta and alpha, during N2 stage. Other examples are shown in Figure S4 – S6 in the supplementary material.

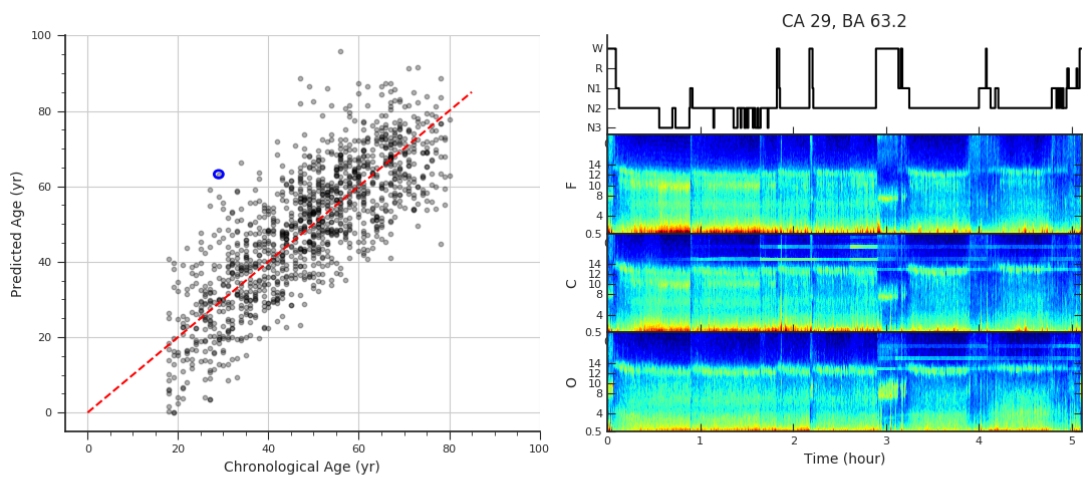

```
[1    theta_kurtosis_C_R]   age norm: -0.00 (-1.94~1.94)  this:  5.33(+)  weight: +1.65  net: 8.82
[2   delta/theta_max_O_N3]  age norm:  0.54 (-1.27~2.35)  this: -1.99(-)  weight: -1.73  net: 4.39
[3    alpha_kurtosis_F_N2]  age norm:  0.67 (-0.76~2.11)  this: -0.56     weight: -3.46  net: 4.26
[4    theta_kurtosis_F_N2]  age norm:  0.70 (-0.76~2.15)  this: -0.95(-)  weight: -2.25  net: 3.70
[5    theta_kurtosis_C_N2]  age norm:  0.69 (-0.75~2.14)  this: -0.78(-)  weight: -2.48  net: 3.66
```

**Figure 6.** A 29-year-old female accumulatively diagnosed as diabetes mellitus, obesity, smoker, cerebrovascular accident, congestive heart failure and acute renal failure not later than 1 year after the PSG recording with higher brain age than chronological age. The blue circle in the top left panel indicates this participant. The top right panel shows the hypnogram and EEG spectrogram from frontal (F), central (C) and occipital (O) channels. The bottom panel shows the top 5 features that, compared to age norm, contribute most positive net effect to the older brain age. The numbers in the parentheses indicates the range of age norm +/-2 standard deviation.

## Discussion

Our study is distinguished by the use of large datasets: 2,621 sleep EEGs from the MGH Sleep Lab, and 3,520 sleep EEGs from the SHHS dataset (Dean et al., 2016; Quan et al., 1997; Redline et al., 1998). The number of EEGs involved in this study is large among relevant "brain age" studies (Cole and Franke, 2017). The large size of the datasets helps to ensure the statistical power as well as less selection bias. A large training set helps ensure that the trained model is not overfit to the particular dataset, improving the ability to generalize when applied beyond the training set. A large testing set allows accurate statistical measurement of how accurately the model predicts brain age.

We use a parametric linear model, which allows easier interpretation by looking at each EEG feature and comparing to the age norm for each feature explicitly. In contrast, nonparametric or semi-parametric models such as kernel methods and Gaussian process models (Franke et al., 2010) estimate age essentially based on the similarity to the training examples, where the contribution of each feature is involved implicitly. An example of interpretability is demonstrated in Figure 6. A notable change in this example is the continual and spreading wide band oscillations. The continual theta power is consistent with prior literature on diabetic encephalopathy, where deficits in glucose metabolism lead to impaired functional connectivity (Baskaran et al., 2013; Franke et al., 2013; Sima, 2010).

The deviations of EEG-based BA from CA might arise from multiple sources. These can be divided into technical and physiological factors. Technical factors include (1) EEG artifacts not removed by preprocessing; (2) constraints of the current model, which assumes brain age is a linear combination of sleep EEG features averaged separately for each sleep stage over the night; (3) imputation of missing sleep stages (Figure S2); and (4) first night effect, where the participants is not used to the sleep lab environment. Neurophysiological factors, our central interest, might arise from (1) night-to-night variability; (2) underlying diagnosed or undiagnosed neurological or psychiatric diseases; (3) general medical health, including diabetes and small vessel cerebrovascular disease; (4) there are substantial genetically-determined stable inter-individual differences in the EEG; (5) exposure to environmental insults. Further studies are needed to clarify the relative contributions of the various types of neurophysiological factors, and to measure the association between EEG-based brain age and various outcomes, such as cognitive performance (Steffener et al., 2016), intelligence (Ujma et al., 2017) and survival (Cole et al., 2017). The deviations suggest that there are unobserved factors, such as diet, exercise and diseases, to further explain the variance. On the other hand, it raises important questions that whether intervention on these factors can change sleep EEG, hence achieving a healthier brain aging process.

It is possible to come up with brain age using multiple modalities, such as EEG, MRI (Cole and Franke, 2017), actigraph, genetic and molecular (Horvath, 2013) biomarkers, which achieves more accurate prediction of brain function and disease risks. Combining features from different modalities help to reduce unobserved variance compared to using single modality. An age norm could also be derived based on multimodal features that have matched chronological age and brain age, thus providing more insights about the aging process. On the other hand, multimodal model sets a higher burden for easy usage or home-based devices since there are more signals to record. Another way to improve performance could be having multiple models for different demographic covariates, such as sex and race.

There are some important limitations of our study. (1) The current dataset does not support the study of night-to-night variability of brain age as well as how it is affected by behaviors. Future studies should measure night-to-night variability, and the degree to which night-to-night differences are found to represent random noise vs. meaningful variation of brain function (e.g. due to differences in sleep quality and/or timing). An important future assessment needs to be longitudinal studies with individual follow-up at multiple time points. (2) We only include ages 18 to 80. The distribution of ages in the dataset is not uniform, where participants in middle age are the most, and participants in young or old age are relatively fewer. This could lead to a systematic bias where old and pathological participants cannot have a much older brain age since the training set does not contain it. (3) Our model is limited by its linear structure. Advanced algorithms such as convolutional neural networks, trained end-to-end from raw EEG data rather than on hand-engineered physiologically-based features as in our model, might be able to predict age more accurately (Biswal et al., 2017; Cole et al., 2016). In addition, whereas we average features within the same sleep stage, representations learned by recurrent neural network models might provide a more powerful summary of the overnight EEG that utilizes more information (e.g. fragmentation of sleep stages) and more accurately reflects brain age.

## Conclusions

Using a machine learning model, the brain age can be inferred from the pattern of brain activity during sleep. Specific diseases and medications, which make the participant deviate from normal aging process, can be reflected from the change in brain age. Using participants with matched chronological age and brain age, an age norm can be developed to provide a direct interpretation of the deviation in aging in terms of EEG features. In summary, the brain age serves as a potential biomarker which sets the stage for EEG-based studies of brain health.

# Acknowledgements

MBW reports grants from NIH-NINDS (NIH-NINDS 1K23NS090900). MTB has received funding from, the Center for Integration of Medicine and Innovative Technology, the Milton Family Foundation, the MGH-MIT Grand Challenge, and the American Sleep Medicine Foundation, and the Department of Neurology. MTB has a patent pending on a home sleep monitoring device, has research agreements with MC10 and Insomnisolv, and consulting agreements with McKesson, International Flavors and Fragrances, and Apple Inc., serves as a medical monitor for Pfizer, and has provided expert testimony in sleep medicine. Dr. Thomas reports the following: 1) Patent, license and royalties from MyCardio, LLC, for an ECG-based method to phenotype sleep quality and sleep apnea; 2) Grant support, license and intellectual property (patent submitted) from DeVilbiss Healthcare; 3) GLG consulting for general sleep medicine; 4) Intellectual Property (patent) for a device using $CO_2$ for central / complex sleep apnea.

This is not an industry supported study, and none of these entities had any role in the study.